\documentclass[11pt]{article}
\usepackage{latexsym,color,amsmath,amssymb, mathtools}
\usepackage{amsthm,graphicx,url,epsfig,algorithmic,textcomp,wasysym,wrapfig,appendix,fancyhdr}

\def\reals{{\mathbb R}}
\def\eps{{\varepsilon}}
\def\A{{\cal A}}
\def\D{{\cal D}}

\def\K{{\cal K}}

\def\S{{S}}
\def\varphiup{ \Phi}

\newtheorem{theorem}{Theorem}[section]

\begin{document}

\title{Computing the Discrete Fr\'echet Distance in Subquadratic Time\thanks{%
Work on this paper by Pankaj Agarwal and Micha Sharir has been
supported by Grant 2006/194 from the U.S.--Israel Binational Science
Foundation. Work by Pankaj Agarwal is also supported by NSF under
grants CNS-05-40347, CCF-06 -35000, IIS-07-13498, and CCF-09-40671,
by ARO grants W911NF-07-1-0376 and W911NF-08-1-0452, by an
NIH grant 1P50-GM-08183-01, and by a DOE grant OEG-P200A070505.
Work by Haim Kaplan has been supported by Grant 2006/204 from the
U.S.--Israel Binational Science Foundation, and by Grant 822/10
from the Israel Science Fund.
Work by Micha Sharir has also been supported
by NSF Grant CCF-08-30272, by Grant 338/09 from the Israel Science Fund,
and by the Hermann Minkowski--MINERVA Center for Geometry at Tel Aviv
University.
Work by Haim Kaplan and Micha Sharir has also been supported by
the Israeli Centers of Research Excellence (I-CORE) program (Center No.~4/11).
Work by Rinat Ben Avraham was supported by the Israel Science Fund Grants
338/09 and 822/10.
  }}

\author{
Pankaj K. Agarwal\thanks{%
  Department of Computer Science, Box 90129, Duke University,
  Durham, NC 27708-0129, USA; {\tt pankaj@cs.duke.edu}}
\and
Rinat Ben Avraham\thanks{%
  School of Computer Science, Tel Aviv University, Tel Aviv 69978,
  Israel; {\tt rinatba@gmail.com}}
\and
Haim Kaplan\thanks{%
  School of Computer Science, Tel Aviv University, Tel Aviv 69978,
  Israel; {\tt haimk@post.tau.ac.il}}
\and
Micha Sharir\thanks{%
  School of Computer Science, Tel Aviv University, Tel~Aviv 69978,
  Israel; and Courant Institute of Mathematical Sciences, New York
  University, New York, NY~~10012,~USA; {\tt michas@post.tau.ac.il }}
}

\maketitle

\begin{abstract}
The Fr\'echet distance is a similarity measure between two curves
$A$ and $B$: Informally, it is the minimum length of a leash
required to connect a dog, constrained to be on $A$, and its owner,
constrained to be on $B$, as they walk without backtracking along
their respective curves from one endpoint to the other. The
advantage of this measure on other measures such as  the Hausdorff
distance is that it takes into account  the ordering of the points
along the curves.

The discrete Fr\'echet distance replaces the dog and its owner by a
pair of frogs that can only reside on $n$ and $m$ specific pebbles
on the curves $A$ and $B$, respectively. These frogs hop from a
pebble to the next without backtracking. The discrete Fr\'echet
distance can be computed by a rather straightforward quadratic
dynamic programming algorithm. However, despite a considerable
amount of work on this problem and its variations, there is no
subquadratic algorithm known, even for approximation versions of the
problem.

In this paper we present a subquadratic algorithm for computing the
discrete Fr\'echet distance between two sequences of points in the
plane, of respective lengths $m\le n$. The algorithm runs in
$O\left(\dfrac{mn\log\log n}{\log n}\right)$ time and uses $O(n+m)$
storage. Our approach uses the geometry of the problem in a subtle
way to encode legal positions of the frogs as states of a finite
automata.
\end{abstract}

\section{Introduction}
\label{sec:introduction}

\paragraph{Problem statement.}
Let $A=(a_1,\ldots,a_m)$ and $B=(b_1,\ldots,b_n)$ be two sequences
of $m$ and $n$ points, respectively, in the plane.
The \emph{discrete Fr\'echet distance} $\delta_{dF}(A,B)$ between $A$
and $B$ is defined as follows. Fix a distance $\delta>0$ and
consider the Cartesian product $A\times B$ as the vertex set
of a directed graph $G_\delta$ whose edge set is
\begin{align*}
E_\delta = & \left\{\Bigr( (a_i,b_j), (a_{i+1},b_j) \Bigl) \: \middle| \,
\|a_i-b_j\|,\;\|a_{i+1}-b_j\| \le\delta \right\} \bigcup \\
& \left\{\Bigr( (a_i,b_j), (a_i,b_{j+1}) \Bigl) \: \middle| \,
\|a_i-b_j\|,\;\|a_i-b_{j+1}\| \le\delta \right\} ;
\end{align*}
here we consider the case where $\|\cdot\|$ is the Euclidean norm.
Then $\delta_{dF}(A,B)$ is the smallest $\delta>0$ for which
$(a_m,b_n)$ is reachable from $(a_1,b_1)$ in $G_\delta$.
Informally, think of $A$ and $B$ as two sequences of stepping stones,
and of two frogs, the $A$-frog and the $B$-frog, where the $A$-frog
has to visit all the $A$-stones in order, and the $B$-frog
has to visit all the $B$-stones in order. The frogs are connected
to each other by a rope of length $\delta$, and are initially
placed at $a_1$ and $b_1$, respectively. At each move, exactly one
of the frogs can jump from its current stone to the next one, which
can be done if and only if its distances to the other frog, before
and after the jump, are both at most $\delta$ (see Figure~\ref{fig:legal_path} for an example of a possible sequence of jumps of the two frogs). Then $\delta_{dF}(A,B)$
is the smallest $\delta>0$ for which there exists a sequence of
jumps that gets the frogs to $a_m$ and $b_n$, respectively. (Note that the frogs cannot backtrack.)

\paragraph{Remark.}
In this formulation we forbid the frogs to jump simultaneously, from a placement $(a_i, b_j)$ to $(a_{i+1}, b_{j+1})$. However, our algorithm can be modified so that it also applies to the variant where such ``diagonal'' moves are also allowed. (See a remark in Section~\ref{subsec:handling_a_block}.)

\paragraph{The continuous Fr\'echet distance.}
The discrete Fr\'echet distance problem is a variant of the (more
standard, continuous) Fr\'echet distance problem. Informally,
consider a person and a dog connected by a leash, each walking along a
path (curve) from its starting point to its end point. Both are
allowed to control their speed, but they cannot backtrack. The
Fr\'echet distance between the two curves is the minimal length of a
leash that is sufficient for traversing both curves in this manner.

More formally, a curve $f \subseteq \mathbb{R}^2$ is a continuous
mapping from $[0,1]$ to $\mathbb{R}^2$. A {\em reparameterization}
is a continuous nondecreasing surjection $\alpha: [0,1]
\rightarrow [0,1]$, such that $\alpha(0)=0$ and $\alpha(1) = 1$.
The Fr\'echet distance $\delta_F(f,g)$ between two curves $f$
and $g$ is then defined as follows:

$$
\delta_F(f, g) = \inf_{\alpha, \beta}\max_{t \in [0,1]} \Bigl\{\|
f(\alpha(t)) -  g(\beta(t)) \| \Bigr\},
$$
where $\|\cdot\|$ is the underlying norm (typically, the
Euclidean norm), and $\alpha$ and $\beta$ are
reparameterizations of $[0,1]$.

\paragraph{The semi-continous Fr\'echet distance.}
One may also consider a hybrid version of the problem, of a person walking a frog. 
Formally, we have a curve $f$ and a sequence $B$ of $n$ stepping stones. We want to find the smallest $\delta > 0$ for which $f$ can be partitioned into $n$ (pairwise openly disjoint) arcs $f_1,\ldots,f_n$
so that the distance of $b_i$ from every point of $f_i$ is at most $\delta$,
for $i=1,\ldots,n$. In this setup, for each $i=1,\ldots,n$, the person walks along $f_i$ 
from its starting point to its end point, while the frog sits at $b_i$. Then, when the person 
reaches the endpoint, the frog jumps to $b_{i+1}$, and they keep moving
this way until all of $f$ and $B$ are traversed.

\paragraph{Remark.}
All three variants of the Fr\'echet distance can be extended, in an obvious manner, to any dimension $d \geq 2$, but in this paper we only consider the planar case.

\paragraph{Background.}
Motivated by a variety of applications, the Fr\'echet distance has been studied extensively in computational geometry
for the past 20 years, as a useful measure for the similarity between
curves~\cite{BPSW05, CDGNW11}.
If data is uniformly sampled, which is often the case in practice, it suffices to compute the discrete Fr\'echet distance between the sequences of vertices of the two curves. 
The extended model that also allows diagonal moves
(as in a preceding remark) can potentially allow us to sample more sparsely along relatively straight portions of the curves.

Eiter and Mannila~\cite{EM94} showed that the discrete Fr\'echet
distance in the plane can be computed
in quadratic time (that is, in $O(mn)$ time).
Later, Aronov et al.~\cite{AHKWW06} have given a
$(1+\eps)$-approximation algorithm which solves the discrete
Fr\'echet distance problem between the vertices of two {\em backbone curves} in near
linear time. Backbone curves are required to have edges whose
lengths are close to 1, and a constant lower bound on the minimal distance between any pair of
vertices; they model, e.g., the backbone chains of proteins. Concerning the continuous Fr\'echet distance problem, Alt
and Godau~\cite{AG95} have shown that the Fr\'echet distance of two
polygonal curves with a total of $n$ edges in the plane can be
computed in $O(n^2 \log n)$ time. A lower bound of $\Omega(n \log
n)$ time for the decision version of the problem, where the task is to
decide whether the Fr\'echet distance between two curves is smaller
than or equal to a given value, was given by Buchin et
al.~\cite{BBKRW07}. They also showed that this bound holds for the
discrete version of the problem as well. It has been an open problem
to compute (exactly) the continuous or discrete Fr\'echet distance 
in subquadratic time. Even the simpler variant, in which we only want to 
solve the decision version of the discrete Fr\'echet distance
problem in the plane in subquadratic
time has still been open. In fact, only a few years ago, Alt~\cite{Alt09} 
has conjectured that the decision subproblem of the (continuous)
Fr\'echet distance problem is 3SUM-hard~\cite{GO95}. 

We note that it is also an open problem to solve the 
approximation versions of the Fr\'echet distance problems in subquadratic time. That is, no 
subquadratic algorithm (in $m$ and $n$, with any reasonable dependence on $\varepsilon$) is known for computing a $(1+\eps)$-approximation
of either variant of the Fr\'echet distance (for arbitrary curves / sequences, with no restrictions on their shape). 

To date, the only subquadratic algorithms known for the Fr\'echet
distance problem (either continuous or discrete) are for restricted
classes of curves, such as the algorithm of Aronov et
al.~\cite{AHKWW06} mentioned above. Other classes of curves
considered so far in the literature include closed convex curves and {\em $\kappa$-bounded}
curves~\cite{AKW04}. A curve is $\kappa$-bounded if, for any pair of
points $a, b$ on the curve, the portion of the curve between $a$ and $b$ is contained in $D(a, \frac{\kappa}{2} \|a-b\|) \cup D(b, \frac{\kappa}{2} \|a-b\|)$, where $D(p, r)$ denotes the disk of radius $r$ centered at $p$.
Alt et al.~\cite{AKW04} showed that the
Fr\'echet distance between two convex curves equals their Hausdorff
distance, and that the Fr\'echet distance between two $\kappa$-bounded
curves is at most $(1+\kappa$) times their Hausdorff distance, and
thus an $O(n \log n)$ algorithm for computing or approximating the Hausdorff distance
(as given in~\cite{Alt09}) can be applied to obtain an efficient exact solution in the convex case or a constant-factor approximation in the $\kappa$-bounded case. Later, Driemel et
al.~\cite{DHW10} provided a $(1 + \eps)$-approximation algorithm for
{\em $c$-packed curves} in $\mathbb{R}^d$ that runs in $O(cn/\eps + cn \log n)$ time, where
a curve $\pi$ is called $c$-packed if the total length of $\pi$
inside any ball is bounded by $c$ times the radius of the ball.

Another variant of the Fr\'echet distance is the {\em weak}
Fr\'echet distance, which, in the person-dog scenario, allows the
person and the dog to also walk backwards. Recently, Har-Peled and
Raichel~\cite{HR11} gave a quadratic algorithm for computing (a
generalization of) the weak Fr\'echet distance between curves. More
specifically, given two simplicial complexes in $\mathbb{R}^d$, and
start and end vertices in each complex, they show how to compute
two curves in these complexes that connect the corresponding start
and end points, such that the weak Fr\'echet distance between these
curves is minimized. Since a polygonal curve is a simplicial
complex, this can be viewed as a generalization of the regular
notion of the weak Fr\'echet distance between curves.

See also~\cite{CDHSW11, DH12} for a few additional results on the Frechet distance.

\paragraph{Our results.}
We present a new algorithm for computing the discrete Fr\'echet distance
whose running time is $O(mn \log \log n / \log n)$ (assuming $m \le n$). 
We first present a procedure for solving the decision version of the problem: Given $\delta > 0$, determine whether 
the discrete Fr\'echet distance between $A$ and $B$ is $\le \delta$. The decision procedure runs in $O(mn \log \log n / \log^2 n)$ time and uses $O(m+n)$ space.
To obtain a solution for the optimization problem,
we combine the decision procedure with a relatively simple explicit binary search, based on a simple procedure for distance selection~\cite{AASS93}. This increases the total running time by only a factor of $O(\log n)$, so
the overall algorithm runs in $O(mn \log \log n / \log n)$ time, which is still subquadratic. Using (a variant of) the procedure in~\cite{AASS93}, the space required by the optimization algorithm remains linear in $m+n$.
The following presentation is therefore mainly focused on the decision procedure, which is the more involved part of our algorithm.

Although not detailed in this abstract, our technique can be extended
so as to compute, within the same time bound, (i) the discrete Fr\'echet
distance between two sequences of points in $\reals^d$, for any $d\ge 3$,
and (ii) the semi-continuous Fr\'echet distance between a sequence of 
points and a curve in the plane. (We do not have at the moment a 
similar extension to the continuous Fr\'echet distance, which is one of the
main open problems raised by our work.)

\paragraph{A brief sketch of the decision procedure.}
Let us first provide a brief description of the decision procedure for a given $\delta >0$. We begin by presenting a slightly less efficient but considerably simpler solution, on which we will then build our improved solution.
Consider the following $0/1$ matrix $M$, whose rows (resp., columns) correspond to the points of $A$ (resp., of $B$).
An entry $M_{i,j}$ of $M$ is equal to 1 if the pair $(a_i, b_j)$ is reachable from the starting placement $(a_1, b_1)$ of the trip with a ``leash'' of length $\delta$. Otherwise, $M_{i,j}$ is equal to 0. In other words, $M_{i,j} = 1$ if the discrete Fr\'echet distance $\delta_{dF}$ between the two prefix subsequences $(a_1,\ldots, a_i)$ and $(b_1,\ldots, b_j)$ is at most $\delta$, and $M_{i,j} = 0$ otherwise. Thus determining the value of $M_{m,n}$ solves the overall decision problem.

$M_{m,n}$ can be obtained by computing all entries of $M$ using dynamic programming, as follows.
If $\|a_1 - b_1\| \le \delta$, we set $M_{1,1} \coloneqq 1$; otherwise, $M_{1,1} \coloneqq 0$ and the decision procedure is aborted right away, since $\delta$ is too small even for the initial placement.
The other elements of the first row of $M$ are then filled in order. Specifically, for each $1 < j \le n$ we set $M_{1,j} \coloneqq 1$ if (a) $M_{1,j-1} = 1$, and (b) 
$\|a_1 - b_j\| \le \delta$; otherwise we set $M_{1, j} \coloneqq 0$. (Clearly, if $M_{1, j} = 0$, for some $0 \le j \le n$, then all the subsequent entries of the first row are also zero.)
Similarly, the first column of $M$ is filled in by setting, for each $1 < i \le m$ in order, $M_{i,1}\coloneqq 1$ if (a) $M_{i-1,1} = 1$, and (b) $\|a_i -b_1\| \le \delta$; otherwise, we set $M_{i, 1} \coloneqq 0$. 
For an arbitrary entry, $1 <i \le m, 1<j \le n$, we set $M_{i,j} \coloneqq 1$ if (a) at least one of $M_{i,j-1}$ and $M_{i-1, j}$ is 1, and (b) 
$\|a_i - b_j\| \le \delta$; otherwise, we set $M_{i,j} := 0$. The cost of this dynamic programming procedure is $O(mn)$.

\begin{figure}[htb]
\begin{center}
\input{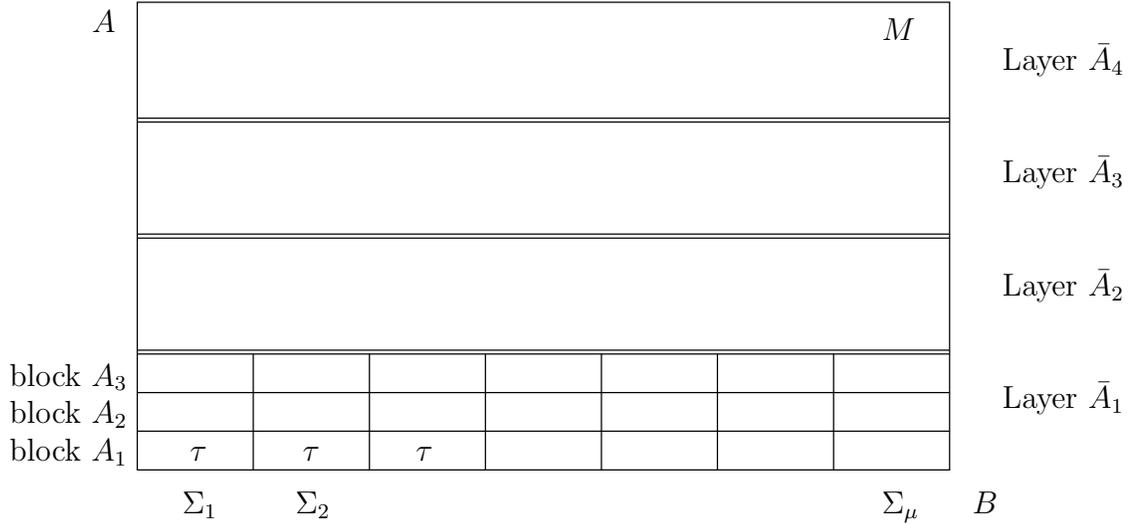}
\caption{\small\sf A partition of $M$ into horizontal strips (and
substrips), which correspond to layers (and blocks) of $A$. $B$ is partitioned into subsequences of length $\tau$. Each subsequence of $B$ corresponds to a single symbol $\Sigma_i$ which the automata $\K^*$ process.}
\label{fig:grid}
\end{center}
\end{figure}

To obtain a subquadratic decision procedure, we cannot compute each value of $M$ explicitly, and instead we only compute certain rows and columns of $M$. 
To be more precise, we partition $A$ into $l = \Theta(m/\log^2 n)$ {\em layers} $A_1, \ldots, A_l$, each of length $c_1\log^2 n$, where $c_1 > 0$ is an appropriate constant 
such that the last point of any layer $A_i$ is the first point of the next layer $A_{i+1}$. We can think of this as a partition of $M$ into $l$ ``horizontal'' strips, each of width $c_1\log^2 n$,
such that the last row of a strip is the first row of the next strip. (See Figure~\ref{fig:grid} for an illustration.)
We then compute, for each strip (in order), the entries of $M$ in the last row of the strip, and we use the values of this row as input for the processing of the next strip.

To obtain the running time bound claimed above, we need to compute the entries of $M$ in each of the $l+1$ ``boundary'' rows (including the first and the last rows) in $O(n \log \log n)$ time. To do so, we further partition each layer $A_i$ into $t = \Theta(\log n)$ {\em blocks}, of length $c_2\log n$ each, where $c_2>0$ is a sufficiently small constant to be specified later (or, alternatively, partition each strip of $M$ into $\Theta(\log n)$ substrips, of width $c_2\log n$ each). As before, the last point of a block is the first point of the next block. We handle each block in $O(n\log \log n/\log n)$ time, using an approach that resembles the execution of a deterministic finite automaton $\K^*$. Somewhat informally, the automaton is constructed from the corresponding block of $A$, and we execute it on a string constructed from the elements of $B$. To achieve the desired running time (in particular, to avoid having to spend $\Theta(n)$ time in ``reading'' the individual elements of $B$), we partition $B$ into $\mu = \Theta(n \log\log n /\log n)$ subsequences of length $\tau = c_3\log n / \log \log n$ each, where $c_3>0$ is yet another constant, and require $\K^*$ to operate on each subsequence, in constant time, as if it were a single symbol.

We note that the compaction of $M$ outlined above is similar to compactions used to solve several related problems. For instance, Baran et al.~\cite{BDP08} present an $o(n^2)$ algorithm for the 3SUM problem on integers of bounded length. (See also algorithms for the {\em edit distance} problem;~\cite{CLZ03, HLLW09, MP80}). However, while the other compactions are purely symbolic, ours is strongly based on the geometry of the problem.
A major difference between our algorithm and the other ones is that in our 
case the input of the problem in itself does not include repetitions 
(that can be used in the compaction). That is, the input points are not 
likely to repeat themselves. We create repetitions artificially by constructing 
the arrangement $\A$ of the disks centered at the points of $A$, and locating 
the points of $B$ in this arrangement. Now, the faces of $\A$ that contain the 
points of $B$ generally repeat themselves. The finite-state automaton $\K^*$
that we construct operates on the faces of $\A$ rather than on the points of $B$,
and this leads to the desired subquadratic performance.
Using such an automaton for the compaction appears to be also a novel
technique.

\paragraph{Organization.}
In Section~\ref{sec:decision_procedure} we describe the decision procedure in detail. In particular, in Section~\ref{subsec:handling_a_block} we show how to deal (slightly less efficiently) with a single block of $A$. We then show, in Section~\ref{subsec:handling_a_layer} how to handle a layer of $A$, which contains $\Theta(\log n)$ such blocks, by combining portions of the processing of the separate blocks into a common procedure that is executed at the layer level. Finally, in Section~\ref{subsec:overall_procedure} we describe the overall decision procedure. (The justification of using blocks of size $\Theta(\log n)$ is deferred to Section~\ref{sec:exp}, where we present a lower-bound construction that indicates that using blocks of larger size may cause our respective automata $\K^*$ to be too large for a subquadratic algorithm.) In Section~\ref{sec:optimization_problem} we show how to combine the decision procedure with an elementary binary search, and obtain the main result of this paper, namely, a subquadratic algorithm for computing the discrete Fr\'echet distance.

\section{The decision procedure}
\label{sec:decision_procedure}
In this section we focus on the decision problem:
Given $\delta>0$, determine whether $\delta_{dF}(A,B) \le \delta$.
By an appropriate scaling, we may assume, without loss of
generality, that $\delta=1$.

As mentioned in the introduction, we partition $A$ into $l = \Theta(m/\log^2 n)$ layers, of size $c_1\log^2 n$ each (where $c_1 > 0$ is an appropriate constant whose value will be set later),
such that the last point of each layer is the first point of the next layer, and process them in order. To process a single layer of $A$, we further partition it into $t = \Theta(\log n)$ blocks, of size $c_2 \log n$ each (where $c_2>0$ is a sufficiently small constant, also to be specified later), such that the last point of each block is the first point of the next block. The algorithm processes the blocks within a layer one by one in order. The purpose of processing a layer is to collect, in a single processing step, information that will be needed by each of its blocks. Then each block is processed separately, in order, and the $M$-entries of its terminal row are computed from those in the initial row, all the way to the terminal row of the entire layer.

\subsection{Handling a single block of $A$} 
\label{subsec:handling_a_block}

Here are the details of processing a single block. To simplify the notation, we denote
the block by $A$; its size $m$ now satisfies
$m=c_2\log n$ (the very last block of the entire sequence may be smaller). Enumerate the points of $A$ as $a_1,\ldots,a_m$.

Regard the points $a_1,\ldots,a_m$ as the centers
of respective unit disks $D_1,\ldots,D_m$, and let $\D$ denote the
sequence of these disks. Consider the arrangement $\A=\A(\D)$ of the
disks, and associate with each face $f$ of $\A$ the subset $\D_f$
of disks containing $f$. For each point $b_i \in B$, denote by $f_i$ the
face of $\A$ containing $b_i$.

\paragraph{Remark.} The description given in this subsection provides the essential ingredients of the processing of a block, but is somewhat lax or vague about precise implementation details, which have to be applied with care to ensure the running time we are after. For example, a naive implementation of the step that finds the faces $f_i$, by $n$ point locations of the points of $B$ in $\A$, is too expensive for our purpose. The layers are used to conglomerate some parts of the processing within their blocks into a single processing step, thereby improving the efficiency of the procedure. More details are provided in the next subsection.

\begin{figure}[htb]
\begin{center}
\input{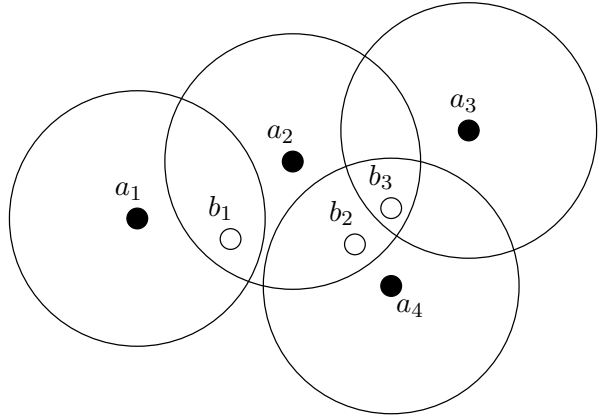}
\caption{\small\sf An illustration of the decision problem of the discrete Fr\'echet distance. The stepping stones of the $A$-frog are the black points. The disks (of radius $\delta$) centered at the points of $A$ form the arrangement $\A$.
The stepping stones of the $B$-frog are the hollow points.
In this example, a legal path of the two frogs is $\bigr((a_1,b_1), (a_2, b_1), (a_2, b_2), (a_2,b_3), (a_3, b_3), (a_4, b_3)\bigl)$.}
\label{fig:legal_path}
\end{center}
\end{figure}

Fix two indices $1\le i\le j\le n$, and call the pair
$\Bigr( (a_1,b_i), (a_m,b_j) \Bigl)$ \emph{valid} if there exists
a path in $G_\delta$ ($G_1$, that is) from $(a_1,b_i)$ to
$(a_m,b_j)$. We can simulate such a path as a sequence of moves
between \emph{basic states}, where each basic state is a pair
$(f,D_k)$, where $f$ is a face of $\A$ and $D_k$ is a disk in $\D_f$.
In each move we either pass from $(f,D_k)$ to $(f',D_k)$, where $f'$
is another face of $\A$ which is also contained in $D_k$, or pass
from $(f,D_k)$ to $(f,D_{k+1})$, if $D_{k+1}$ also belongs to $\D_f$
(i.e., also contains $f$). See Figure~\ref{fig:legal_path}. In the original problem (involving the complete unpartitioned $A$) we would have to start at
$(f_1,D_1)$ and
to reach $(f_n,D_m)$ (now with $m$ equal to the original size of $A$), using a sequence of legal moves between basic states, of the types just described, that corresponds to a path in $G_1$ from $(a_1, b_1)$ to $(a_m, b_n)$. (For this, though, we would need to construct the huge arrangement of the disks for the entire sequence $A$, which would have been far too expensive.)
In the refined version we start at $(f_i,D_1)$ and have to reach $(f_j,D_m)$ along a similar sequence of moves, for arbitrary indices $i \le j$ (and for the much smaller size $m$ of a block).
This represents the situation
where the portion of the trip of the $B$-frog that corresponds to
the passage of the $A$-frog through the points of the present block
$A$ starts at $b_i$ and ends at $b_j$.

Note that, in view of this
interpretation, we are only interested in placements $(a_1,b_i)$ that
can be reached (through the preceding blocks of the complete $A$-sequence) 
from the starting placement of the whole trip. We refer to such a placement 
as a {\em reachable} position of the frogs.
Let the flag $\varphi_i = \varphi(b_i)$ indicate whether the placement
$(a_1,b_i)$ is reachable through 
the preceding blocks of $A$ (in which case $\varphi_i = 1$), or not ($\varphi_i = 0)$; in this notation we hide the dependence of $\varphi_i$ on the preceding layers and blocks.
Note that if $A$ is the first block, then $\varphi_i = 0$ for each $i>1$, 
since $(a_1,b_i)$ is not reachable through the (empty set of)
preceding blocks. For $i=1$ we set $\varphi_1 =1$ if $b_1 \in D_1$,
and otherwise abort the entire procedure, since the frogs lie at their 
starting positions at distance $> 1$.

We can store the data maintained by this process in more compact form. 
To do so, we define an
\emph{aggregate state} (to which we refer as a \emph{state} for short)
to be a pair $(f, \S_f)$, where $\S_f$ is a subset of $\D_f$; we refer
to $\S_f$ as the set of valid disks (associated with our state).
The set $\S_f$ is assumed to have the property, dictated by the transition
rules for the frogs, that if $D_k\in \S_f$
and $D_{k+1}\in\D_f$ then $D_{k+1}$ also belongs to $\S_f$.

A state $(f,\S_f)$ and a pair $(g, \varphi)$, where $g$ is a
face of $\A$, and $\varphi$ is a binary flag, determine a 
transition into a new state $(g, \S_g)$, where
$\S_g\subseteq \D_g$ consists of those disks $D_k\in\D_g$
for which there exists $j\le k$ such that (i) $D_j\in \S_f$,
and (ii) the entire run $D_j,D_{j+1},\ldots,D_k$ is contained
in $\D_g$. Furthermore, if $\varphi = 1$, then $\S_g$ also contains the 
maximal prefix of disks in $\D$ (starting with $D_1$) that is contained
in $\D_g$.
The passage from $(f,\S_f)$ to $(g,\S_g)$ is called a
\emph{valid transition}.

The interpretation of this setup is as follows. The state $(f,\S_f)$
signifies that (a) the $B$-frog is now at a point that belongs to $f$,
and the $A$-frog lies at the center $a_k$ of some disk $D_k\in \S_f$,
and (b) this position has been reached via a legal sequence of
interweaving $A$-moves and $B$-moves, starting from $(a_1,b_1)$ 
(if $A$ is the first block of the whole sequence),
or from some placement $(a_1,b_i)$ (if $A$ is an intermediate block),
which is reachable from the starting positions of the frogs (so $\varphi_i$ is 1).
Moreover, for the specific sequences of stepping stones for the
$A$-frog and the $B$-frog, the $A$-frog cannot lie at the center $a_k$ of any
disk $D_k\notin \S_f$.

The valid transition from $(f,\S_f)$ to
$(g,\S_g)$ means that, for any disk $D_k\in \S_g$, we can get the
$A$-frog to lie at its center $a_k$, and get the $B$-frog to lie in
$g$, by taking a disk $D_j$ as in the definition of the valid
transition, assuming that the $A$-frog lies at $a_j$ and the $B$-frog
lies in $f$ (in accordance with the above interpretation of $(f, S_f)$), moving the $B$-frog to $g$ (which is possible since $D_j$
also belongs to $\D_g$), and then moving the $A$-frog through the
centers $a_{j+1},\ldots,a_k$, all at distance at most $1$ from the
$B$-frog (or, if $j=k$, let the $A$-frog stay put). Moreover, if the last move of the
$B$-frog is from $f$ to $g$, and the $A$-frog lies at the center of
some disk in $\S_f$, then the centers of the disks in $\S_g$ are the
only possible locations that the $A$-frog can reach (with this single
hop of the $B$-frog).

In addition, the flag $\varphi$ allows the $B$-frog to
appear ``out of nowhere'' in the middle of the first row of the block, 
in case a position $(a_1, b_i)$, where $b_i \in g$, is reachable from
the starting placement of the whole trip.
This means that we can get the $A$-frog to lie at $a_1$, and the $B$-frog to 
lie in $g$, by some path starting at the starting position of the entire trip of the frogs, 
and moves solely through the points of the preceding blocks of the full sequence $A$ (once the $B$-frog has reached $g$, the $A$-frog can move through the centers of the disks in the prefix of $\D_g$ contained in $\S_g$, and stop at any of these centers before the $B$-frog makes its next move).
 
The compression of basic states into aggregate states resembles
the construction of a deterministic finite automaton (DFA) from
a nondeterministic finite automaton (NFA). 
This is not accidental; we have already hinted that the algorithm simulates the moves of such an automaton, and
the resemblance will become more relevant as we continue to present
the algorithm.

\noindent{\bf Remark:}
If we want to also
consider the variant where the frogs are allowed to jump simultaneously
from a placement $(a_i,b_j)$ to $(a_{i+1},b_{j+1})$ (provided that $\|a_i-b_j\| \le 1$ and $\|a_{i+1}-b_{j+1}\| \le 1$), we only need to modify the above rules of a valid transition.
Specifically, a state $(f,\S_f)$ and a pair $(g, \varphi)$, where $g$ is a
face of $\A$, and $\varphi$ is a binary flag, determine a 
transition into a new state $(g, \S^\prime_g)$, where $\S^\prime_g\subseteq \D_g$
is the union of $\S_g$, as defined above, and of another set $\bar{\S}_g \subseteq \D_g$, consisting of those disks $D_k\in\D_g$
for which there exists $j\le k$ such that (i) $D_j\in \S_f$,
and (ii) the entire run $D_{j+1},\ldots,D_k$ is contained
in $\D_g$. (so the disk $D_j$ is not required to belong to the run).

\paragraph{A DFA interpretation.}
We can interpret the setup just described
as a construction of 
a deterministic finite automaton $\K$, as follows; for the convenience of the reader, we include the following short glossary of the main notations used in this construction.
\\ \\
$f$ or $f_i$ (or $g$) - a face of $\A(\D)$.\\
$F$ - string of $n$ faces.\\
$F_k$ - a substring of $F$ (of length $\tau$).\\ \\
$\varphi$ or $\varphi_i$ -  a binary flag.\\
$\Phi$ - string of $n$ flags.\\
$\Phi_k$ - a substring of $\Phi$ (of length $\tau$).\\ \\
$\sigma$ - a pair $(f, \varphi)$ of a face $f$ and a flag $\varphi$.\\
$\Sigma$ - string of $n$ pairs. \\
$\Sigma_k$ - a substring of $\Sigma$ (containing $\tau$ pairs).\\

The states of $\K$ are the
aggregate states $(f,\S_f)$, where $f$ is a face of the corresponding
disk arrangement $\A$ and $\S_f\subseteq \D_f$.
The $i$-th `character' in the string that $\K$ has to process is the pair $(g_i, \varphi_i)$, where $g_i$ 
is the face of $\A$ that contains $b_i$, and 
$\varphi_i$ is a flag indicating whether $(a_1, b_i)$ is a reachable position of the two frogs (in the sense defined above, with respect to the whole trip).
The transition from a state $(f,\S_f)$ on reading
the pair $(g_i, \varphi_i)$ is to $(g_i,\S_{g_i})$, where $\S_{g_i}$ is defined as above.
The string that $\K$ has to process to handle the current block $A$
is thus the string of pairs
$\Sigma = \Bigl( (f_2,\varphi_2),\ldots, (f_n,\varphi_n) \Bigr)$,
where $f_2,\ldots, f_n$ are
the (not necessarily neighboring)
faces of $\A$ containing the corresponding actual points $b_2,\ldots,b_n$ of the $B$-sequence, and
$\varphi_2,\ldots, \varphi_n$ are
the respective flags associated with $b_2,\ldots,b_n$, as defined above.

The starting state of $\K$ is the
state $(f_1,\S_{f_1})$, where $f_1$ is the face containing $b_1$,
and where $\S_{f_1} = \emptyset$ if 
$\varphi(b_1)=0$, or $\S_{f_1}$ is the largest prefix of $\D$ 
contained in $\D_{f_1}$ if $\varphi(b_1)=1$.
Note that in the construction of $\K$ we are not given
that prefix --- $\K$ is defined in terms of $A$ only.
Furthermore, when $\K$ does read the $B$-string $\Sigma$ and reaches a state $(f_i, \S_{f_i})$ it outputs a new flag 
$\varphi(b_i)$, which is $1$ if
$D_m\in \S_{f_i}$ and is $0$ otherwise. The points $b_i$ with
$\varphi(b_i) = 1$ are exactly those for which
$(a_m,b_i)$ is reachable in $G_1$ (from the beginning of the whole trip).
In this context, we can think of $\K$ as a {\em Moore machine}~\cite{Moore56} --- a finite-state transducer that associates an output value with each state. We can thus associate the output flag $\varphi(b_i)$ with the state $(f_i, \S_{f_i})$.
The output flag $\varphi(b_i)$ will be used later, as an input for the next block (see Section~\ref{subsec:overall_procedure}).

As noted earlier, if $A$ is the first block of the whole sequence, each flag of the input sequence $\Sigma$, except for the first one, is equal to zero. For the first position $(a_1, b_1)$ of the first block, we assume that $b_1 \in D_1$  ;
otherwise, as already mentioned, we abort the decision procedure right away, reporting 
that the Fr\'echet distance $\delta_{dF}(A,B)$ is greater than $1$.
We thus set, after verifying this constraint, $\varphi_1 = \varphi(b_1) = 1$.

\paragraph{Remark.}
The automaton $\K$ is constructed from the block $A$ only, without knowing anything about the sequence $B$. Consequently, for each face $f$ of the arrangement $\A$, we need to prepare states $(f, \S_f)$ for each subset $\S_f \subseteq \D_f$ that might arise via some sequence of stepping stones of the $B$-frog. As shown in Section~\ref{sec:exp}, there are situations where the number of such feasible subsets may be exponential in $|\D_f|$ (that is, in $m$). This is why we need to take $m = c_2\log n$, with $c_2$ sufficiently small, to control the size of $\K$ and the time needed to construct it (so that they are both sublinear in $n$).

\paragraph{Constructing an efficient DFA.}
To obtain an overall procedure with subquadratic running time, we modify the construction of $\K$ to obtain a 
somewhat more efficient automaton $\K^*$ to handle a block $A$.
There are two major improvements in the construction of $\K^*$. The first, whose detailed description is deferred to Section~\ref{subsec:handling_a_layer}, is to construct $\K^*$ in terms of the finer arrangement $\A^*$ of the disks centered at all the $\Theta(\log^2 n)$ points of the $A$-sequence within the layer containing the current block. Informally, the reason for doing it (explained in detail in Section~\ref{subsec:handling_a_layer}) is that it saves us the need to locate the $B$-points in each of the coarser block arrangements, a process that would be too expensive for our purpose. Nevertheless, so as not to throw at the reader the two improvements at the same time, we present here the construction of $\K^*$ solely in terms of the current block arrangement $\A$ and then modify it in the next subsection.

The second improvement aims to allow $\K^*$ to process the $B$-dependent string $\Sigma$ in a faster manner. Specifically, we
modify $\K$ so that each input character that it reads is a 
string of $c_3\log n/ \log \log n$ consecutive input characters of $\Sigma$, where $c_3>0$ is a sufficiently small constant, whose value will be determined later.
That is, we partition the input string $\Sigma$
of $\K$ into 
$\mu = \Theta(\frac{n \log \log n}{\log n})$ substrings $\Sigma_1, \Sigma_2, \ldots, \Sigma_\mu$ of size 
$\tau=c_3\log n /\log \log n$ each; the last substring may be shorter.
The states of $\K^*$ are the
same aggregate states $(f,\S_f)$ of $\K$.
When $\K^*$ is at state $(f, \S_f)$ and is given a substring 
$\Sigma_k = ((f_1,\varphi_1),\ldots, (f_\tau,\varphi_\tau))$, 
it moves to state $(f_\tau,\S_{f_\tau})$, where $(f_\tau,\S_{f_\tau})$
is the state that $\K$ would have reached from $(f, \S_f)$ after processing the input substring 
$\Sigma_k$ character by character. (The subscripts used in the enumeration of the pairs of $\Sigma_k$ start at 1 for the sake of simplicity. This involves a slight abuse of notation, because $(f_1, \varphi_1)$ denotes here the first pair of $\Sigma_k$ and not the first pair of the entire string $\Sigma$.)

Furthermore, a transition of $\K^*$ from a state $(f, \S_f)$ to a state 
$(f_\tau,\S_{f_\tau})$ as above, produces an output string 
$\varphiup_k = (\varphi_1,\ldots, \varphi_\tau)$, where $\varphi_j$ 
is the output of $\K$ when it reaches the state $(f_j,\S_{f_j})$ (again, under the new enumeration convention). Recall that we regarded $\K$ as a Moore machine, where the output flags $\varphi_j$ are associated with the corresponding states $(f_j,\S_{f_j})$.
However, here the state $(f_\tau,\S_{f_\tau})$ that $\K^*$ moves to after reading $((f_1,\varphi_1),\ldots, (f_\tau,\varphi_\tau))$ cannot determine by itself the output string 
$\varphiup_k$, which requires knowledge of the full sequence $((f_1,\varphi_1),\ldots, (f_\tau,\varphi_\tau))$ that led $\K^*$ to $(f_\tau,\S_{f_\tau})$. More specifically, the flags comprising $\varphiup_k$ are determined by the states $(f_j,\S_{f_j})$ that $\K$ traversed on the way to $(f_\tau,\S_{f_\tau})$. To avoid having to look at each intermediate state $(f_j, S_{f_j})$ separately, we observe that all these states are implicitly encoded in the transition {\em edge} of $\K^*$ that takes us from $(f, S_f)$ to $(f_\tau, S_{f_\tau})$ upon reading $\Sigma_{k}$. We can therefore regard $\K^*$ as a {\em Mealy machine}~\cite{Mealy55} ---
a finite-state transducer that associates an output value with each transition edge.

The rest of the description of $\K^*$ remains the same as that of $\K$.

In the following, we describe how to construct $\K^*$ so that a state 
transition can be carried out in constant time. (A full description of the construction of $\K^*$ will be given in the next subsection.) As is shown later,  
executing a transition of $\K^*$ in constant time is  
essential for obtaining the subquadratic running time of the whole optimization procedure.

To construct $\K^*$, we build the transition table $T$, according to the rules stated above. Since $T$ is constructed independently of the input string $\Sigma$, we must prepare, for each state $(f, \S_f)$ of $\K^*$, all possible transitions to a new state.
That is, given a state $(f, \S_f)$ we store, for each possible input substring $\Sigma_k$ of length $\tau$, the state $(g, S_g)$ that $\K$ moves to after processing $\Sigma_k$ (assuming that $\K$ was in state $(f, \S_f)$ just before reading $\Sigma_k$).
To be more precise, we prepare the transition table $T$ of $\K^*$ as a collection of arrays $L_{(f, \S_f)}$, one array for each state $(f, \S_f)$ of $\K^*$. The array $L_{(f, \S_f)}$ of a fixed state $(f, S_f)$ is defined so that, for each index $j$ encoding a substring $\Sigma_k$ (details of the encoding are provided next), $L_{(f, \S_f)}[j]$ is the pair $((g,\S_g), \varphiup_{k})$, where $(g, \S_g)$ is the state that $\K^*$
moves to after processing $\Sigma_k$ (assuming that $\K$ was in state $(f, \S_f)$ just before reading $\Sigma_k$), and $\varphiup_{k}$ is the 
output substring of flags that corresponds to this transition.

To complete the description of $T$, we now describe a simple encoding scheme that converts each string $\Sigma_k = \left((f_1,\varphi_1), \ldots, (f_\tau, \varphi_\tau)\right)$ of $\tau$ pairs into an integer $e(\Sigma_k)$ of $O(\log n)$ bits. 
To do so, each face $f$ of the arrangement of the disks of $A$ is given an integer label $e(f)$ in the range $(0,\ldots, c\log^4 n)$, for an appropriate absolute constant $c$. (We will later explain, as part of the full description of the construction of $\K^*$, why we use this range and how to generate the labels efficiently.) Clearly, at most $\beta = \log(c \log^4 n) =  c^\prime \log \log n$ bits are needed for such a label, for another absolute constant $c^\prime$ (close to $4$).
We now put 
\begin{equation}
\label{eqn:e_F}
e(\Sigma_k) = \displaystyle\sum_{i=1}^\tau e(f_i)\cdot 2^{\beta(i-1)+\tau} + \displaystyle\sum_{i=1}^\tau \varphi_i \cdot 2^{i-1},
\end{equation}
and note that $e(\Sigma_k)$ does indeed consist of only $\tau(\beta+1) = O(\log n)$ bits. Clearly, this is a one-to-one encoding.

With this setup, each state transition 
can be executed in constant time.
Specifically, when $\K^*$ is in state $(f, \S_f)$ and is given the encoding $e(\Sigma_k)$ of an 
input substring $\Sigma_k$, we follow a pointer to the array $L_{(f, \S_f)}$ and retrieve its entry $L_{(f, \S_f)}[e(\Sigma_k)]$ in constant time. This gives us the next state $(g, \S_g)$ and the corresponding output bitstring $\varphiup_{k}$. Hence, the
execution of $\K^*$, when given $O\left(\frac{n \log \log n}{\log n}\right)$ substrings as above, takes $O\left(\frac{n \log \log n}{\log n}\right)$ time.
This cost excludes the computation of the indices $e(\Sigma_k)$, which will be discussed in the next subsection.

The size of (number of entries in) $T$ is the number 
of states of $\K^*$, multiplied by
the number of possible input substrings for $\K^*$.
The latter number is $2^{(\beta +1)\tau} \le 2^{c^{\prime\prime}\log n}$, where $c^{\prime\prime}$ is proportional to $c_3$, which we choose sufficiently small so as to have $c^{\prime\prime} < 1/4$, say. The number of states of $\K^*$ is $O\left(m^2 2^m\right)$, where $m = c_2\log n$ is the size of a block:\footnote{The first improvement in $\K^*$, deferred to Section~\ref{subsec:handling_a_layer}, will cause the number of states to increase to $O(m^4 2^m)$, which will have negligible effect on the performance of the algorithm; see below for details.} There are $O(m^2)$ faces in the disk arrangement, and, in view of the construction given in Section~\ref{sec:exp}, we use the pessimistic bound of $2^m$ on the number of possible subsets $\S_f$ for any fixed face $f$. Choosing $c_2$ sufficiently small, we can ensure that the number of states of $\K^*$ is at most $O(n^{1/4})$, say. Hence the size of $T$ is $O(n^{1/2}) = O(n \log \log n / \log n)$, and it can be built within the same asymptotic time bound.


\subsection{Handling a layer of $A$} 
\label{subsec:handling_a_layer}
In order to make the whole procedure efficient, we need to construct quickly the encodings of the input strings for the automata of the blocks of $A$. Note that we cannot even afford linear (i.e., $O(n)$) time for this preparation for each block, because this would result in the overall bound $O(mn/\log n)$ for the running time of the decision procedure, which, multiplied by the number $O(\log n)$ of binary search steps, would yield $O(mn)$ overall running time, which defeats our goal of obtaining a subquadratic solution.

This is the reason for using a two-stage partitioning of $A$, first into layers of size $c_1\log^2 n$ each, and then into blocks of size $c_2\log n$ each --- The preparation of the strings is done mainly at the layer level, thereby making the cost sublinear for each block.

Here are the details of this preprocessing step. Fix a layer $\bar{A}$ of $A$, which contains $t = \Theta(\log n)$ blocks, of size $c_2\log n$ each, which we enumerate as $A_1, \ldots, A_t$. As before, the last point of $A_i$, for $1 \le i < t$, is the first point of $A_{i+1}$. We process $A_1, \ldots,A_t$ in order, in much the same way as described in Section~\ref{subsec:handling_a_block}, except that some of the preparatory steps are grouped together, and take place during the preprocessing of $\bar{A}$.
 
In more detail, we first construct the arrangement 
$\bar{\A} = \bar{\A}(\bar{\D})$, where $\bar{\D}$ is the set
of $c_1\log^2 n$ unit disks centered at the points of $\bar{A}$; the number of faces of $\bar{\A}$ is at most $c\log^4 n$, for an appropriate constant $c$ (the same constant appearing in the encoding in the previous subsection). 
We preprocess $\bar{\A}$ for
efficient point location, using any of the standard
techniques, in $O(\log^4 n \log\log n)$ time.
Fix a block $A_j$ of $\bar{A}$, and note that each face $f$ of $\bar{\A}$ is a subface of a face $f^{(j)}$ of the arrangement $\A_j$ of the disks centered at the points of $A_j$. We find these correspondences by preprocessing each $\A_j$ for fast point location, and then, for each face $f$ of $\bar{\A}$ we pick an arbitrary point in $f$ and locate it in $\A_j$, thereby obtaining $f^{(j)}$. In this way each face of $\bar{\A}$ stores $t$ pointers to its ``super-faces'' $f^{(j)}$, for $j = 1,\ldots,t$. 

Next, for each point $b_i$ of the $B$-sequence, we locate the face $f_i$ of $\bar{\A}$ 
containing $b_i$, using the point location structure. 
This takes $O(n \log\log n)$ time. We obtain a sequence $F = (f_1,f_2,\ldots, f_n)$ of faces of $\bar{\A}$, and we partition it into $\mu$ subsequences $F_1, \ldots, F_\mu$, each consisting of $\tau$ consecutive faces, where $\mu = \Theta(n \log \log n/ \log n)$ and $\tau = c_3\log n/\log\log n$, as in the preceding subsection.

Now comes the other improvement in the construction of the block-automata $\K^*$ considered in Section~\ref{subsec:handling_a_block}. Specifically, since the number of faces of $\bar{\A}$ is at most $c\log^4 n$, we label each face $f$ of $\bar{\A}$ by an integer $e(f)$ in the range $(0,\ldots,c\log^4 n)$. For each of the $\mu$ subsequences $F_k$ of $F$, say $F_k = (f_1,\ldots,f_\tau)$, we compute the ``partial'' index (cf. (\ref{eqn:e_F}))

\begin{equation}
e_0(F_k) = \displaystyle\sum_{i=1}^\tau e(f_i)\cdot 2^{\beta(i-1)+\tau}.
\end{equation}

Note that, given the labels $e(f_i)$, $e_0(F_k)$ can be computed by $O(\tau)$ additions and multiplications (or, rather, shifts).
In addition, note that this index is common to all the blocks of $\bar{A}$; we stress again that each such partial index is computed only once within the layer $\bar{A}$. 

Now fix a block $A_j$ of $\bar{A}$, and consider the construction of its automaton $\K^*_j$.
Except for the fact that the faces of $\bar{\A}$ that we use here are smaller than the respective faces of the block arrangement $\A_j$, the states $(f, S_f)$ and the transition rules for $\K^*_j$ are very similar to those used in Subsection~\ref{subsec:handling_a_block}. More specifically, each face $f_0$ of $\A_j$ is now the union of some faces of $\bar{\A}$. Every state of the form $(f_0, S_{f_0})$ that we had before is now copied, for each face $f \subseteq f_0$ of $\bar{\A}$, to a state $(f, S_{f_0})$. A similar copying is applied to the transition rules. That is, consider first the non-compacted automaton $\K_j$. If it is at a state $(f, S_f)$ and reads a pair $(g, \varphi)$, where $f$ and $g$ are now faces of $\bar{\A}$, we apply the same transition rule that the original $\K_j$ obeys when it is at state $(f_0, S_f)$ and reads $(g_0, \varphi)$, where $f_0$ (resp., $g_0$) is the face of $\A_j$ containing $f$ (resp., $g$). We now obtain the new version of $\K^*_j$ from the new version of $\K_j$ in the same manner as above. That is, when $\K^*_j$ is at state $(f, S_f)$ and reads a substring $\Sigma_k = ((f_1,\varphi_1),\ldots,(f_\tau, \varphi_\tau))$ of $\Sigma$, where now $f, f_1, \ldots, f_\tau$ are faces of $\bar{\A}$, it moves to the state $(f_\tau, S_{f_\tau})$ obtained by running the new $\K_j$ on the pairs of $\Sigma_k$ one by one.

The total time for computing the $\mu$ indices $e_0(F_k)$ is linear in $n$. This is tolerable since we carry out this computation only once for the entire layer $\bar{A}$. 
However, each of the subsequences $\Sigma_k$ that we feed into the various block automata $\K^*_j$ has a second ``component'' that depends on the input flags at the first row of the respective block $A_j$. Specifically, each $\Sigma_k$ is of the form $((f_1,\varphi_1),\ldots,(f_\tau, \varphi_\tau))$, which we can represent by the pair $(F_k, \Phi_k)$, where $F_k = (f_1,\ldots,f_\tau)$ and $\Phi_k = (\varphi_1,\ldots,\varphi_\tau)$.
The subsequences $F_k$ are computed once, at the layer level, and do not change from block to block, but the subsequences $\Phi_k$ do depend on the blocks. In terms of the encoding in (\ref{eqn:e_F}) we have 
\begin{equation}\label{eqn:e_sigma_k}
e(\Sigma_k) = e_0(F_k) + e_0(\Phi_k),
\end{equation}
where
\begin{equation}\label{eqn:e_Phi_k}
e_0(\Phi_k) = \sum_{i=1}^\tau \varphi_i\cdot 2^{i-1}
\end{equation}
is simply the bitstring consisting of the flags in $\Phi_k$.

We can easily construct the automata $\K^*_j$ in such a way that the output of each transition is the encoding $e_0(\Phi_k)$ of the corresponding sequence $\Phi_k$. 
Assuming that this is the case, we process a block $A_j$ as follows.
Let $\Phi_1,\ldots, \Phi_\mu$ denote the output flag subsequences from
the execution of the preceding automaton $\K^*_{j-1}$ (or from the
execution of the last automaton in the preceding layer, or from the
initialization of the entire procedure). By assumption, we are
actually given the encodings $e_0(\Phi_1),\ldots,e_0(\Phi_\mu)$ (the
computation of these bitstrings during initialization is trivial and
inexpensive), and we substitute them in (\ref{eqn:e_sigma_k}) to
obtain $e(\Sigma_1),\ldots,e(\Sigma_\mu)$. This computation takes
$O(\mu) = O(n \log\log n/\log n)$ time for each block, for a total of
$O(n \log\log n)$ time for the whole layer. We now run (the modified
automaton) $\K^*_j$ on the string $(e(\Sigma_1),\ldots,e(\Sigma_\mu))$
and obtain the output sequence $e_0(\Phi_1^\prime),\ldots,
e_0(\Phi_\mu^\prime))$, where $\Phi_1^\prime,\ldots,\Phi_\mu^\prime$
are the flag subsequences output by the state transitions of $\K^*_j$,
which are the input for the next automaton.

The analysis in Section~\ref{subsec:handling_a_block} shows that, with an appropriate choice of the constants $c_1, c_2, c_3$, the construction of the automata $\K^*_j$, for $j = 1,\ldots, t$, takes a total of $O(n)$ time (in fact, much smaller if we so wish).
Processing a single block 
costs $O(n\log \log n / \log n)$ time (see
Section~\ref{subsec:handling_a_block} and the preceding paragraph).
Since $\bar{A}$ contains $\Theta(\log n)$ blocks,\footnote{This step
in the analysis is the reason for restricting the size of a layer to
$\Theta(\log^2 n)$ points of $A$, that is, to $\Theta(\log n)$ blocks.} the total cost for processing $\bar{A}$ is $O(n \log \log n)$. (This includes the cost of the point location stage within $\bar{\A}$, which is also $O(n \log \log n)$.) In conclusion, processing a single layer, including the processing of each of its blocks, takes a total of $O(n \log \log n)$ time.

The space required for this procedure is linear in $n$, since we need to store the subsequences of faces of $\bar{\A}$, which are used as input for each $\K^*_i$.
The space used for handling a block $A_i$ of $\bar{A}$ is sub-linear in $n$ (see Section~\ref{subsec:handling_a_block}), and can be freed after processing $A_i$. Hence, the total space required for processing $\bar{A}$ is still linear in $n$.

\subsection{The overall procedure}
\label{subsec:overall_procedure}
To obtain an overall algorithm with subquadratic time, we partition the original sequence $A$ into $\Theta(m/\log^2 n)$ layers
$\bar{A}_1,\bar{A}_2,\ldots$, each (except possibly for the last one) consisting of
$c_1\log^2 n$ points,
and so that the last point of $\bar{A}_i$ is the first point of $\bar{A}_{i+1}$
for each $i$. We then process $\bar{A}_1, \bar{A}_2, \ldots$ in succession.

To process a layer $\bar{A}_i$, we use the procedure of Section~\ref{subsec:handling_a_layer}. If $\bar{A}_i$ is not the last layer of $A$, we use the output sequence $\varphiup_1,\ldots, \varphiup_\mu$ of $\bar{A}_i$ as input for $\bar{A}_{i+1}$ (as described in Section~\ref{subsec:handling_a_layer}). Otherwise, $\bar{A}_i$ is the last layer of $A$, and we use the last flag $\varphi_\tau$ of the last subsequence $\varphiup_\mu$ to determine the outcome of the decision process --- if $\varphi_\tau=1$ we report that
$\delta_{dF}(A,B)\le \delta$; otherwise $\delta_{dF}(A,B) > \delta$.

Processing each layer $\bar{A}_i$ of $A$ takes $O(n\log \log n)$ 
time, so processing the $\Theta(m/\log^2 n)$
layers, takes $O(mn \log \log n/ \log^2 n)$ time.
The space required for handling a layer $\bar{A}_i$ of $A$ is linear in $n$ (see Section~\ref{subsec:handling_a_layer}), and it can be freed after handling $\bar{A}_i$. Hence, the space required by the decision procedure is only $O(n + m)$ (we need $O(m)$ space for storing $A$). 

Hence, we obtain the following intermediate result.

\begin{theorem} \label{th:decision}
Given two sequences $A$, $B$ of stepping stones, of respective
sizes $m$ and $n$, with $m \le n$, and a parameter $\delta >0$, we can decide, using $O\left(\frac{m n \log \log n}{\log^2 n}\right)$ time and $O(n+m)$ space, whether $\delta_{dF}(A,B) \le \delta$.
\end{theorem}

\paragraph{Remark.}
The above procedure determines whether $\delta_{dF}(A, B) > \delta$
or $\delta_{dF}(A, B) \le \delta$.
In the latter situation, there is no need to discriminate between $\delta_{dF}(A, B) < \delta$ and $\delta_{dF}(A, B) = \delta$, since this could easily be done upon termination of the binary search, as described in Section~\ref{sec:optimization_problem}, by comparing 
two consecutive critical values of $\delta$ reached at the end of the search. See Section~\ref{sec:optimization_problem} for more details.

\section{Solving the optimization problem}
\label{sec:optimization_problem}
We use the decision procedure in Section~\ref{sec:decision_procedure} to solve 
the optimization problem, as follows.
First note that the critical values of $\delta$, in which an edge is added to the
graph $G_\delta$ (as $\delta$ increases), are the pairwise distances 
between a point of $A$ and a point of $B$. Hence, it suffices to perform 
a binary search over all possible $m n$ such distances,
and execute the decision procedure in each step of the search.
At each such step, the corresponding pairwise distance is the $l$-th smallest pairwise distance in $A \times B$ for some value of $l$. 
We can find this distance, e.g., using a variant of one of the algorithms of Agarwal et al.~\cite{AASS93}, which runs in time close to $O(n^{3/2})$. This algorithm can easily be adapted to the ``bichromatic'' scenario, where we consider only distances between the pairs in $A \times B$ (as opposed to finding distances between the points of a single set).
More specifically, we use a variant of the simpler (sequential) decision procedure of~\cite{AASS93}. We partition the set $A$ into $\lceil m/n^{1/2} \rceil$ smaller subsets, each of size at most $n^{1/2}$, and operate on each subset independently, coupled with the whole $B$. In processing such a subset $A_i$, we construct the arrangement of the disks of radius $\delta$ centered at the points of $A_i$, and locate the points of $B$ in this arrangement, exactly as in~\cite{AASS93}. Altogether, this yields the number of pairs in $A \times B$ at distance at most $\delta$, which is what the decision procedure needs. The overall cost of this procedure is $O(n^{3/2}\log n)$. Finally, we solve the optimization version of the distance selection algorithm using parametric searching, increasing the running time to $O(n^{3/2}\log^3 n)$. This running time is subsumed by the cost of the decision procedure of Section~\ref{sec:decision_procedure}.\footnote{Although there are more efficient algorithms for distance selection, which run in close to $O(n^{4/3})$ time~\cite{AASS93, KS97}, this simple-minded solution suffices for our purpose, and it has the advantage that it only uses linear storage.}

Since we call the decision procedure $O(\log n)$ times during the search, we obtain the following main result of the paper.

\begin{theorem} \label{th:optimization}
Given two sequences $A$, $B$ of stepping stones, of respective
sizes $m$ and $n$, with $m \le n$, we can compute the discrete Fr\'echet distance 
between $A$ and $B$ in ${\displaystyle O\left(\frac{m n \log \log n}{\log n}\right)}$ time and $O(n+m)$ space.
\end{theorem}

\section{An exponential lower bound on the number of states}
\label{sec:exp}

An interesting question that pops up right away in the design of the algorithm is how large
can $\K^*$ be. That is, how many aggregate states (and transition rules) can one have.
Unfortunately, the following construction shows that this number can be exponential
in $m$ in the worst case.

\begin{figure}[htb]
\begin{center}
\input{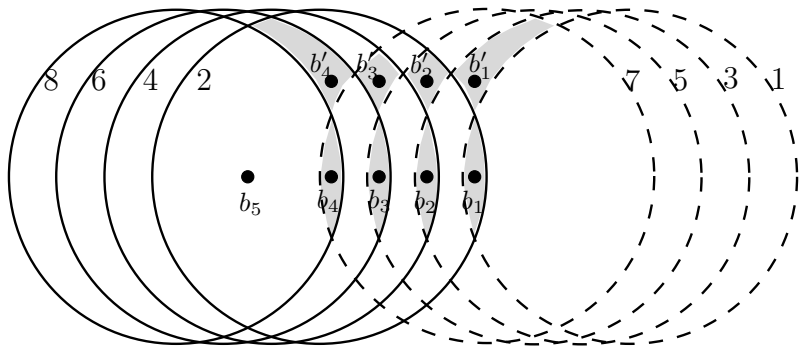}
\caption{\small\sf A configuration of disks with an exponential number of states. The red disks are drawn solid and the blue disks are drawn dashed.}
\label{fig:explow}
\end{center}
\end{figure}

The construction, depicted in Figure~\ref{fig:explow}, uses an even number of disks; with a slight
abuse of notation, we denote their number by $2m$.
Enumerate the disks as $D_1,D_2,\ldots,D_{2m}$ and their
respective centers as $a_1,a_2,\ldots,a_{2m}$.
All these centers lie on the $x$-axis in the right-to-left order
$a_1,a_3,\ldots,a_{2m-1},a_2,a_4,\ldots,a_{2m}$.
The centers of the even-indexed disks (red disks for short)
are sufficiently close to each other, so that these disks have
a large common intersection. The odd-indexed disks (blue disks
for short) are placed so that, for each $k=1,\ldots,m$,
$D_{2k-1}$ intersects $D_{2k}$ (in a small cap) but is disjoint
from $D_{2k+2}$ (the second condition is vacuous for $k=m$).

We next place $2m+1$ points
$b_1,b'_1,b_2,b'_2,\ldots,b_m,b'_m,b_{m+1}$ (or, rather,
select $2m+1$ corresponding faces
$f_1,f'_1,f_2,f'_2,\ldots,f_m,f'_m,f_{m+1}$ of the
resulting arrangement of the $2m$ disks).
For each $i=1,\ldots,m$, we take $f_i$ to be the cap
$D_{2i-1}\cap D_{2i}$ (by construction, and as shown in the figure, these are
indeed faces of the arrangement).
We take $f'_i$ to be the face lying directly above $f_i$, so
that in order to go from $f_i$ to $f'_i$ we need to exit the
two disks $D_{2i-1}$ and $D_{2i}$ (and not to cross the
boundary of any other disk). Finally, we take $f_{m+1}$ to
be the intersection face of all the red (even-indexed) disks.

We regard $(a_1,b_1)$ as the starting position of the frogs,
where $b_1$ is any point in $f_1$ and
$a_1$ is the center of $D_1$, and the goal position is
$(a_{2m},b_{m+1})$, where $b_{m+1}$ is any point in $f_{m+1}$
and $a_{2m}$ is the center of $D_{2m}$.

By construction, $\D_{f_{m+1}}$ consists of all the $m$ red
disks. We claim that for every subset $\S\subseteq \D_{f_{m+1}}$,
$(f_{m+1},\S)$ is a valid state, obtaining the asserted exponential
number of states. To be more precise, the claim is that for any
such $\S$ we can construct a sequence $B=B_\S$ of points, which (i) starts
at $b_1$ and ends at $b_{m+1}$, (ii) contains all the points $b_1, b_2,\ldots, b_{m+1}$ (in this order), and (iii) contains some of the points $b'_1, \ldots, b'_m$, so that if it contains $b'_j$ then $b'_j$ appears between $b_j$ and $b_{j+1}$. The sequence $B_\S$ has the property that for any $D\in \S$, as
the $B$-frog moves through the sequence $B_\S$, the $A$-frog can
execute a sequence of corresponding moves, so that it reaches
at the end the center of $D$, and this cannot be achieved (for the same sequence $B_\S$) for
any $D\notin \S$. For simplicity, we only specify the sequence
of faces of $\A$ containing the points of $B$, rather than the
points themselves (although the figure depicts the points too).

So let $\S \subseteq \D_{f_{m+1}}$ be given. We associate with $\S$ the following
sequence $F_\S$ of faces. We start with the subsequence
$(f_1,f_2,\ldots,f_m,f_{m+1})$ and, for each $D_{2k}$ \emph{not in}
$\S$, we insert $f'_k$ into $F_\S$, between $f_k$ and $f_{k+1}$.
Figuratively, the corresponding sequence $B_\S$, which proceeds
from right to left, is a mixture of sharp vertical detours
(corresponding to red disks not in $\S$) and of short horizontal
moves (for red disks in $\S$).

We next argue that $F_\S$ does indeed generate the state
$(f_{m+1},\S)$. Consider a red disk $D_{2k}$ not in $\S$. When
the $B$-frog follows the detour from $f_k$ to $f'_k$ and
then to $f_{k+1}$, it leaves $D_{2k-1}$ and $D_{2k}$ and then
re-enters $D_{2k}$ (and $D_{2k+2}$).
The maximal run of disks which ends at $D_{2k}$ and is contained in 
$\D_{f_{k+1}}$, includes $D_{2k}$ only, since $D_{2k-1} \notin \D_{f_{k+1}}$. In 
addition, $D_{2k}$ does not belong to $\D_{f'_k}$, so in particular 
$D_{2k} \notin \S_{f'_k}$. Hence, $D_{2k} \notin \S_{f_{k+1}}$, 
because there is no 
valid transition (in this setup) from $(f'_k, \S_{f'_k})$ to $(f_{k+1}, \S_{f_{k+1}})$ such that $D_{2k} \in \S_{f_{k+1}}$ (see the
 rules for a valid transition in Section~\ref{subsec:handling_a_block}).
From this point on, 
the path is fully outside of $D_{2k-1}$, 
so, as easily verified by induction, $D_{2k}$ will not appear in any of the following states, including the state $(f_{m+1}, \S_{f_{m+1}})$, as claimed.
(The reader might wish to interpret this argument in terms
of the actual moves of the frogs.)

Consider next a red disk $D_{2k}$ that belongs to $\S$. It
suffices to show that when the $B$-frog reaches $f_k$, the
$A$-frog could have executed a sequence of preceding moves that
gets it to the center of $D_{2k}$; this is because, 
from this point on, the $B$-frog remains inside $D_{2k}$ (note that,
by construction, we do not execute the detour via $f'_k$), so the
$A$-frog simply has to stay put at the center of $D_{2k}$
and wait for the end of the sequence of moves of the $B$-frog. 

Note that $f_1$ is contained in all blue disks and in $D_2$.
In particular, this implies the asserted property for $k=1$:
The $A$-frog goes from the center of $D_1$ to the center of
$D_2$ before the $B$-frog moves, and stays there till the end.
In general, $f_j$ is contained in the blue disks
$D_{2j-1},D_{2j+1},\ldots,D_{2m-1}$ and in the red disks
$D_2,D_4,\ldots,D_{2j}$. What the $A$-frog needs to do is
to ensure that, for each $j<k$, it lies at the center of
$D_{2j+1}$ by the time the $B$-frog gets to $f_j$. This is
easily argued by induction on $j$. The $A$-frog can do this
for $j=1$, because $f_1$ lies in $D_1,D_2,D_3$. For larger
values of $j$, assume that the $A$-frog is at the center of
$D_{2j-1}$ when the $B$-frog is at $f_{j-1}$. If the path goes
straight to $f_j$, it exits $D_{2j-3}$ and then enters $D_{2j}$.
Since the $A$-frog is at the center of $D_{2j-1}$, it can now
move to the center of $D_{2j}$ and then to the center of
$D_{2j+1}$, as desired. If the path goes to $f_j$ via $f'_{j-1}$,
it exits $D_{2j-3}$ and $D_{2j-2}$, then re-enters $D_{2j-2}$
and then enters $D_{2j}$. However, since the $A$-frog is already
at the center of $D_{2j-1}$, these additional exit and re-entry
are irrelevant for it, and it can now move to the center of
$D_{2j+1}$ as above. Finally, when the $B$-frog moves to $f_k$,
the $A$-frog, which is now at the center of $D_{2k-1}$, moves
to the center of $D_{2k}$ and stays there.
This completes the argument.

\paragraph{Remark.}
It is a challenging open problem to circumvent this exponential lower bound on the number of possible states. Of course, we have exponentially many states because of the existence of exponentially many possible $B$-sequences. Is it possible, for example, to reduce the number of states significantly by some sort of examination of the specific input $B$-sequence? As already remarked, the existence of potentially exponentially many states is the major bottleneck for the efficiency of the algorithm.
In the same vein, it would be interesting to find properties of the sequences $A$, $B$
that guarantee that the number of aggregate states is much smaller. In a sense,
this would hopefully subsume (so far, for the discrete and semi-continuous cases only)
the earlier studies involving special classes of curves and/or sequences~\cite{AKW04, AHKWW06, DHW10}.

\section{Discussion and open problems}
We obtained an algorithm for computing the discrete Fr\'echet distance
between two sets of points, which runs in subquadratic time. A natural
open problem that arises right away is whether this algorithm can be
extended to compute the continuous Fr\'echet distance between two
polygonal curves in subquadratic time. Even solving the
semi-continuous Fr\'echet distance problem in subquadratic time might
be interesting at this point. It is also interesting to know if this
time bound, which is still rather close to quadratic, can be further
reduced (see the remark at the end of the preceding section).

\end{document}